\documentclass[letterpaper,12pt]{article}
\usepackage{amssymb}

\usepackage{ifpdf}
\ifpdf
	\usepackage[pdftex]{graphicx}
	\usepackage[pdftex,unicode,implicit]{hyperref}

	\hypersetup{
  	pdftitle     = {On the consistency of the Ho\v{r}ava Theory},
  	pdfkeywords  = {},
  	pdfauthor    = {J. Bellor\'{\i}n and A. Restuccia},
  	pdfcreator   = {pdf\LaTeXe\ with package \flqq hyperref\frqq},
  	pdfproducer  = {pdf\LaTeXe\ with package \flqq hyperref\frqq},
  	pdfpagemode  = UseNone,  %change this to FullScreen if you want...
  	pdffitwindow = true,  % Open the pdf file without the sidebars open.
  	unicode      = true,
  	plainpages   = true,
  	colorlinks   = true,  % links have no different colors..
  	citecolor    = blue,  
  	urlcolor     = blue,
  	linkcolor    = blue
	}

	\newcommand{\arXiv}[1]{
		{\href{http://www.arXiv.org/abs/#1}{arXiv:#1}}}

\else
  \usepackage[dvips]{graphicx}
  \usepackage[unicode,implicit]{hyperref}

  \newcommand{\arXiv}[1]{arXiv:#1}
  
\fi

\begin{document}

\begin{flushright}
{\small
SB/F/379-10\\
Feb $27^{\rm th}$, $2012$}
%\normalsize
\end{flushright}
\vspace*{10mm}

\begin{center}

\vspace{1ex}
{\bf \LARGE On the consistency of the Ho\v{r}ava Theory}
\vspace{5ex}

%authors
{\sl\large Jorge Bellor\'{\i}n}$\,^{a,}$\footnote{\tt jorgebellorin@usb.ve}
{\sl\large and Alvaro Restuccia}$\,^{a,b,}$\footnote{\tt arestu@usb.ve}
\vspace{3ex}

$^a${\it Departamento de F\'{\i}sica, Universidad Sim\'on Bol\'{\i}var, Valle de Sartenejas,\\ 
1080-A Caracas, Venezuela.} \\[1ex]
$^b${\it Department of Physics, Universidad de Antofagasta, Chile.}

\vspace*{5ex}
{\bf Abstract}
\begin{quotation}{\small
With the goal of giving evidence for the theoretical consistency of the Ho\v{r}ava Theory, we perform a Hamiltonian analysis on a classical model suitable for analyzing its effective dynamics at large distances. The model is the lowest-order truncation of the Ho\v{r}ava Theory with the detailed-balance condition. We consider the pure gravitational theory without matter sources. The model has the same potential term of general relativity, but the kinetic term is modified by the inclusion of an arbitrary coupling constant $\lambda$. Since this constant breaks the general covariance under space-time diffeomorphisms, it is believed that arbitrary values of $\lambda$ deviate the model from general relativity. We show that this model is not a deviation at all, instead it is completely equivalent to general relativity in a particular partial gauge fixing for it. In doing this, we clarify the role of a second-class constraint of the model.
}\end{quotation}

\end{center}

%\newpage

\vspace*{5ex}

%%%%%%%%%%%%%%%%%%%%%%%%%%%%%%%%%%%%%%%%%%%%%%%%%%%%%%%%%%%%%%%%%%%%%%%%%%%
%%%%%%%%%%%%%%%%%%%%%%%%%%%%%%%%%%%%%%%%%%%%%%%%%%%%%%%%%%%%%%%%%%%%%%%%%%%
%%%%%%%%%%%%%%%%%%%%%%%%%%%%%%%%%%%%%%%%%%%%%%%%%%%%%%%%%%%%%%%%%%%%%%%
%%%%%%%%%%%%%%%%%%%%%%%%%%%%%%%%%%%%%%%%%%%%%%%%%%%%%%%%%%%%%%%%%%%%%%%

There have been a lot of interest about Ho\v{r}ava's proposal of a new theory of gravity which in principle has a renormalizable quantum version \cite{Horava:2009uw} (an important part of the conceptual and technical basis was previously developed in Ref.~\cite{Horava:2008ih}). To build such a theory, Ho\v{r}ava has proposed to abandon the principle of space-time relativity as a fundamental symmetry of nature, reducing the freedom to perform coordinate transformations to those transformations that preserve some preferred universal time-like foliation. The advantage of this scheme is that one can include higher spatial-derivative terms in the Lagrangian that render the theory renormalizable. According to Ho\v{r}ava's point of view, Einstein's general relativity would be recovered in some way (at first sight nonevident) as a large-distance approximation of the complete, nonrelativistic theory.

The complete Ho\v{r}ava theory has an extra physical degree of freedom that is not present in general relativity. It is commonly assumed that this is an unavoidable consequence of its ``smaller'' gauge symmetry group and henceforth any model possesing the same foliation-preserving symmetry has the extra degree of freedom. However, the counting of the physical degrees of freedom of a given model is an issue that cannot be guessed \emph{a priori} only on the basis of the gauge symmetries. After all, gauge symmetries are just convenient redundances in the variables used to define a system. Rigorous computations must be performed by, for example, using the canonical formalism. 
%The extra mode of the complete Ho\v{r}ava theory would not be in conflict with the %aim of having an UV completion of general relativity whenever it decouples in some %way at large distances.

Some recent works have been devoted to the issue of the extra mode in Ho\v{r}ava Gravity. This has been focused in Ref.~\cite{Blas:2009yd} from the side of the Lagrangian equations of motion and in Refs.~\cite{Li:2009bg, Henneaux:2009zb,Pons:2010ke} using the canonical formalism. The authors of Ref.~\cite{Li:2009bg} suggested that there are not second-class constraints in the theory and there is no any physical degree of freedom left by the constraints. In Ref.~\cite{Blas:2009yd} an extra scalar mode, with only one field as Cauchy data, was found with an apparent strong coupling at the extremely low IR. Perturbative computations has been done in several works, in particular in Refs.~\cite{Horava:2009uw,Horava:2008ih,Blas:2009yd,Charmousis:2009tc,Gong:2010xp,Park:2009hg}. Among these papers, a detailed analysis of linear perturbations around a cosmological background using the Hamiltonian scheme has been done in Ref.~\cite{Gong:2010xp}. Those authors have found the presence of only two physical degrees of freedom in the linear-order perturbation of the pure Ho\v{r}ava gravity, a result that agrees with Ref.~\cite{Park:2009hg}.

In this paper we take the classical model obtained from the truncation of the potential at lowest order in curvature of the nonprojectable Ho\v{r}ava theory with detailed balance. We show that this model, which we shall call the $\lambda R$-model, is completely equivalent to Einstein's general relativity in certain partial gauge fixing for it, including the fact that the model has only two propagating degrees of freedom.

Truncated theories are of interest since the operators of low order in curvature become dominant as the distance scale grows. For the largest distances (deep IR) one can forget about high-order operators that are important rather for quantum effects, and keep only the $R$ term in the potential (second-order in derivatives), as in Einstein's general relativity. Therefore, the equivalence we find between the $\lambda R$-model and general relativity is quite relevant since it is precisely at large distances where one expect general relativity to be recovered.

The crucial step in our approach is to realize that the condition emerging when the preservation of the Hamiltonian constraint is imposed is not a condition for the lapse function at all, instead it is a second class constraint that restricts another scalar of the set of canonical variables -the trace of the canonical momentum\footnote{In Ref.\cite{Kocharyan:2009te} it was suggested that non-relativistic theories of gravity with a space-time dependant lapse function are typically inconsistent due to the mere presence of this condition.}. As a subsequent step, one is forced to preserve in time this second class constraint and this leads to an equation that is the one that really allows to determine the lapse function as a Lagrange multiplier. At the end one has two physical degrees of freedom and the constraints are the generator of the spatial diffeomorphisms and two second class constraints, the Hamiltonian and the trace of the momentum. In general relativity the Hamiltonian constraint is also present but it is of first-class behavior. It is well know that, under certain circumstances, a theory with two second class constraints can be given in an equivalent formulation with an enhanced gauge symmetry. For this, one of the second class constraints is casted as the first class constraint generating the extra symmetry and the orther one is interpreted as a chosen gauge fixing condition for the extra symmetry. On the basis of this, we show by exact computations that this kind of relation precisely holds for the $\lambda R$-model and general relativity.

The condition we have mentioned was also studied in Ref.~\cite{Henneaux:2009zb} for the same $\lambda R$-model, but regarding it as an equation for the lapse function. Those authors arrived at the conclusion that the only solution for the lapse function must be vanishing in all points. This scenario would be catastrophic for the Ho\v{r}ava Theory since the Lagrangian is singular when the lapse function is zero. They also mention that there are points of the phase space, called by them nongeneric points, that avoids the vanishing of the lapse function. One of the nongeneric points they mention has the vanishing of the trace of the momentum, $\pi = 0$. They indicated that this yields general relativity in a particular gauge. We are going to show that the \emph{only} solution for the $\lambda R$-model is exactly $\pi = 0$, no extra conditions are required, the lapse function is not zero and it is determined by a consistent elliptic equation. The $\lambda R$-model is then completely consistent on itself and equivalent to general relativity.

Let us start with the exact computations. The action of the $\lambda R$-model\footnote{We consider only the theory without the projectability condition, where the lapse function has a general dependence on space and time, since this is the scenario that leads directly to general relativity. The choice of a projectable lapse function as a definitory condition leads to a different theory that cannot be smoothly deduced from the nonprojectable case, as we are going to see in our model. The Hamiltonian analysis for the counterpart of the action (\ref{horavaaction1}) in the projectable case has been done in Ref.~\cite{Kobakhidze:2009zr}.} has the same form of the ADM action of general relativity but with the inclusion of an arbitrary coupling constant $\lambda$,
\begin{equation}
S = 
\int dt d^3 x \sqrt{g} N ( G^{ijkl} K_{ij} K_{kl}  + R ) \,,
\label{horavaaction1}
\end{equation}
where $R$ is the spatial Ricci scalar and the kinetic variable $K_{ij}$ is the extrinsic curvature defined by
\begin{equation}
 K_{ij} = 
 \frac{1}{2N} ( \dot{g}_{ij} - 2 \nabla_{(i} N_{j)} ) \,.
\label{k}
\end{equation}
Of course, this requires that $ N\neq 0$ in all space-time. This restriction avoids singularities in the Lagrangian, a requirement that also arises in the ADM approach of general relativity. $G^{ijkl}$ is a generalization of the De Witt metric (a four-indices metric),
\begin{equation}
G^{ijkl} = 
\frac{1}{2} (g^{ik} g^{jl} + g^{il} g^{jk} )
- \lambda g^{ij} g^{kl} \,,
\end{equation}
With the value $\lambda = 1$ the De Witt metric is recovered. For the case of $\lambda \neq {1}/{3}$ the inverse of $G^{ijkl}$ is given by
\begin{equation}
\mathcal{G}_{ijkl} = 
\frac{1}{2} (g_{ik} g_{jl} + g_{il} g_{jk} )
- \frac{\lambda}{3\lambda - 1} g_{ij} g_{kl} \,.
\label{inverseg}
\end{equation}
Ho\v{r}ava included the constant $\lambda$ based upon the fact that the two contributions for the kinetic term are separately invariant under the gauge symmetries of the theory. Following the logic of effective theories, they emerge as independent contributions of quantum corrections \cite{Horava:2008ih}. Due to the presence of $\lambda$, which is in general different from one, this model can be naively considered as a minimal deviation from general relativity. This lore is supported by the fact that the inclusion of $\lambda$ breaks the full space-time diffeomorphisms symmetry of the theory. However, this does not imply \emph{a priori} that both theories are physically inequivalent.

The conjugate momenta of $N$ and $N_i$ vanish automatically for the Lagrangian given in Eq.~(\ref{horavaaction1}). As usual, we eliminate those momenta from the phase space and consider $N$ and $N_i$ as Lagrange multipliers instead of canonical variables. The momentum $\pi^{ij}$ conjugated to $g_{ij}$ is found to be
\begin{equation}
\frac{\pi^{ij}}{\sqrt{g}} = 
G^{ijkl} K_{kl} \,.
\label{pi}
\end{equation}

We make reasonable assumptions about the boundary conditions: the spatial manifold is noncompact, at the infinite boundary the spatial metric $g_{ij}$ goes to a finite, nonsingular metric, $N$ goes to a nonzero value and $\pi^{ij}$ goes to zero. To be more specific, in asymptotically flat coordinates it is required that as $r$ approaches to infinity \cite{Regge:1974zd},
\begin{equation}
 \begin{array}{rclrcl}
 g_{ij} & = & \delta_{ij} + \mathcal{O}(1/r) \,, 
\hspace{2em}
 \pi^{ij} & = & \mathcal{O}(1/r^2) \,,
\\[2ex]
 N & = & 1 + \mathcal{O}(1/r) \,,
\hspace*{2em}
 N_i & = & \mathcal{O}(1/r) \,.
\end{array}
\label{asymptotics}
\end{equation}

From the relation (\ref{pi}) it is evident that the variable $\dot{g}_{ij}$ can be explicitly solved in terms of the momentum $\pi^{ij}$ only when the metric $G^{ijkl}$ is invertible. For this reason we are forced to consider separately the Hamiltonian analysis for the cases $\lambda \neq \frac{1}{3}$ and $\lambda = \frac{1}{3}$. 

%%%%%%%%%%%%%%%%%%%%%%%%%%%%%%%%%%%%%%%%%%%%%%%%%%%%%%%%%%%%%%%%%%%%%%
%%%%%%%%%%%%%%%%%%%%%%%%%%%%%%%%%%%%%%%%%%%%%%%%%%%%%%%%%%%%%%%%%%%%%%
%%%%%%%%%%%%%%%%%%%%%%%%%%%%%%%%%%%%%%%%%%%%%%%%%%%%%%%%%%%%%%%%%%%%%%
%%%%%%%%%%%%%%%%%%%%%%%%%%%%%%%%%%%%%%%%%%%%%%%%%%%%%%%%%%%%%%%%%%%%%%
%%%%%%%%%%%%%%%%%%%%%%%%%%%%%%%%%%%%%%%%%%%%%%%%%%%%%%%%%%%%%%%%%%%%%%

\paragraph{Hamiltonian analysis for $\lambda \neq {1}/{3}$ \\} 
It is easy to solve Eq.~(\ref{pi}) for $K_{ij}$ and $\dot{g}_{ij}$. After doing this,
we perform the Legendre transformation to obtain the Hamiltonian,
\begin{equation}
\begin{array}{rcl}
H & = &
\left< N \mathcal{H} + N_i \mathcal{H}^i \right > \,,
\\[2ex]
\mathcal{H} & \equiv &
\mathcal{G}_{ijkl} {\displaystyle\frac{\pi^{ij} \pi^{kl}}{\sqrt{g}}} 
 - \sqrt{g} R \,,
\\[2ex]
\mathcal{H}^i & \equiv &
- 2 \nabla_j \pi^{ji} 
= -2 ( \partial_j \pi^{ji} + \Gamma_{jk}{}^i \pi^{jk} ) \,,
\end{array}
\label{hamiltonian}
\end{equation}
where we use the brackets $\left<{\:}\right>$ to denote integration over all the spatial manifold, $\left< \mathcal{F} \right> \equiv \int \mathcal{F} d^3 x$, and the standard definition for covariant derivatives on densities is understood. Notice that this Hamiltonian has been deduced under the assumption $N \neq 0$, hence we must keep this restriction in the Hamiltonian in order to be able to associate it to the Ho\v{r}ava Theory.

Now we must apply Dirac's algorithm to extract all the constraints of the theory and to ensure their preservation in time. The primary constraints are obtained by taking variations w.r.t.~$N$ and $N_i$, which yields $\mathcal{H}$ and $\mathcal{H}^i$ as the primary constraints of the theory. To continue we must demand the preservation in time of them. Since $\mathcal{H}^i$ is the generator of spatial diffeomorphisms, it is clear that the Poisson brackets of $\mathcal{H}^i$ with $\mathcal{H}$ and itself correspond to infinitesimal coordinate transformations on them. Hence these brackets are proportional to $\mathcal{H}$, $\mathcal{H}^ i$ and their derivatives, yielding no new constraints (see Ref.~\cite{DeWitt:1967yk}). More interesting is the preservation in time of $\mathcal{H}$, which, by the same argument, requires to compute only the bracket of $\mathcal{H}$ with itself. By straightforward computations we obtain
\begin{equation}
\{ \left< \epsilon \mathcal{H} \right> , \left< \eta \mathcal{H} \right>\} = 
2 \int d^3 y \epsilon \pi^{ij} \left[
  \nabla_i \nabla_j \eta - \left(\frac{\lambda - 1}{3\lambda - 1}\right)
    g_{ij} \nabla^2 \eta \right]
- (\epsilon \leftrightarrow \eta) \,.
\label{bracketHH}
\end{equation}
Now the bracket we are interested in is obtained by setting $\epsilon = \delta^3 (\vec{x} - \vec{y})$ and $\eta = N$. This yields
\begin{equation}
\{ \mathcal{H} , \left< N \mathcal{H} \right> \} = 
 2 \left(\frac{\lambda - 1}{3\lambda - 1}\right) 
 \left( N \nabla^2 \pi
   + 2 \partial_i N \nabla^i \pi \right)
+ 2 \mathcal{H}^i \partial_i N + N \nabla_i \mathcal{H}^i \,,
\label{brackethh}
\end{equation}
where $\pi \equiv g_{ij} \pi^{ij}$. For $\lambda = 1$ we get the expected result of general relativity: The Hamiltonian constraint $\mathcal{H}$ is of first-class behavior and no more constraints are generated. However, to ensure the preservation in time of the Hamiltonian constraint for $\lambda \neq 1$ we must impose a new condition,
\begin{equation}
N \nabla^2 \pi + 2 \partial_i N \nabla^i \pi
= 0 \,,	
\end{equation}
which can be simplified to the equation
\begin{equation}
\nabla_i \left( N^2 \nabla^i \pi \right)
= 0 \,.
\label{preconstraint}
\end{equation}

Eq.~(\ref{preconstraint}) is the condition we mentioned in the Introduction. We may multiply the Eq.~(\ref{preconstraint}) by $\pi/ \sqrt{g}$ and then integrate it over all the spatial manifold. This yields, after integration by parts,
\begin{equation}
\int \frac{d^3 x}{\sqrt{g}} {N^2 \left(\nabla_i \pi \right)^2}
= 0 \,.
\end{equation}
Since $N^2 / \sqrt{g} > 0$, this equation is equivalent to the equation
\begin{equation}
\nabla_i \pi \equiv \sqrt{g} \partial_i \left(\frac{\pi}{\sqrt{g}}\right) 
= 0 \,.
\end{equation}
The general solution to this equation is that the combination ${\pi}/{\sqrt{g}}$ is an arbitrary function of time, but the spatial-boundary condition $\pi^{ij}|_{\infty} = 0$ forces this function to be zero. Therefore, we get that the condition (\ref{preconstraint}) is totally equivalent to the constraint
\begin{equation}
\pi = 0 \,.
\label{picero}
\end{equation}

It is easy to see that the scalar-density $\pi$ is the generator of the infinitesimal conformal transformations on $g_{ij}$ and $\pi^{ij}$,
\begin{equation}
 \{ g_{ij} , \left< \epsilon \pi \right> \}  = 
    \epsilon g_{ij} \,,
\hspace*{2em}
  \{ \pi^{ij} , \left< \epsilon \pi \right> \} = 
    - \epsilon \pi^{ij} \,,
\end{equation}
being the weight of the momenta $\pi^{ij}$ under these transformations the opposite of $g_{ij}$. However, conformal transformations are not symmetries of the $\lambda R$-model (nor of general relativity), thus one expects that constraint $\pi$ is a second-class constraint.

We also remark that, although in the $\lambda R$-model one starts with the assumption that $\lambda$ is an undetermined coupling constant, once the constraint $\pi$ is imposed $\lambda$ drops out completely from the Hamiltonian, thus at the end one sees that the value of $\lambda$ is completely irrelevant for this model.

Now we must impose the preservation in time of the new constraint $\pi$. Following the same arguments mentioned above about the role of $\mathcal{H}^i$, we may see that the important bracket to compute is $\{ \pi , \left< N \mathcal{H} \right> \}$. It is straightforward to obtain
\begin{equation}
\{ \left< \epsilon \pi \right> , \left< \eta \mathcal{H} \right> \}
=
\int d^3 y \epsilon \left[ - 2 \sqrt{g} ( \nabla^2 \eta - R \eta )
 + \frac{3}{2} \eta \mathcal{H} \right] \,,
\label{bracketpiH}
\end{equation}
hence
\begin{equation}
\{ \pi , \left< N \mathcal{H} \right> \}
=
- 2 \sqrt{g} ( \nabla^2 - R ) N
 + \frac{3}{2} N \mathcal{H} \,.
\end{equation}
Therefore, the preservation in time of the constraint $\pi$ leads to the following equation for $N$:
\begin{equation}
( \nabla^2 - R ) N \cong 0 \,.
\label{nequation}
\end{equation}
This is a strongly elliptic equation for $N$, there always exists a solution of it with the correct asymptotic behavior. Indeed, if for large $r$ one makes the asymptotic perturbation $ N = 1 + n$, where $n$ is of infinitesimal order, then Eq.~(\ref{nequation}) is compatible with the expected asymptotic behavior of the variables $g_{ij}$ and $n$ given in Eqs.~(\ref{asymptotics}). Dirac's algorithm ends with Eq.~(\ref{nequation}).

The equation (\ref{nequation}) does not allow to impose, for general spatial configurations, the so called projectability condition, which consists of imposing $N$ to be a function only of time. Therefore, we see that for theories as the $\lambda R$-model and the Ho\v{r}ava theory the projectability condition cannot be regarded as a smooth limit of the general theory. On the other hand, by imposing the projectability condition at the beginning as a definitory condition one gets the global Hamiltonian constraint $\left<\mathcal{H}\right> = 0$ instead of the local one. Since in the theory with projectability $\left< \mathcal{H} \right>$ is a first-class constraint, its conservation in time is ensured. No more constraints are needed in this case. Due to the lacking of the local Hamiltonian constraint, it is presumably that general relativity is not recovered in the theory with projectability condition\footnote{A perturbative analysis of the theory with projectability has been done in Ref.~\cite{Bogdanos:2009uj}}.

Let us summarize our results on (local) constraints and also count the number of physical degrees of freedom of the $\lambda R$-model. The canonical variables are $g_{ij}$ and $\pi^{ij}$, which sum up for twelve real variables. $\mathcal{H}^i$ is a first-class constraint (three constraints). It is easy to see that the r.h.s. of the Eq.~(\ref{bracketHH}) vanishes in the constrained phase space, such that $\mathcal{H}$ commutes with itself and also $\pi$ commutes with itself. However, from Eq.~(\ref{bracketpiH}) we conclude that $\mathcal{H}$ does not commute with $\pi$ in the constrained phase space, hence both are the second class constraints of the theory (two constraints). The number of physical degrees of freedom in the canonical space is given by
\begin{equation}
\begin{array}{l}
\mbox{(\# Physical D.O.F.)} =  
\nonumber \\[2ex]
 \hspace*{2em}  \mbox{(\# Canonical var.)} 
   - 2 \times \mbox{(\# 1$^{\mbox{\tiny st}}$ class const.)} 
     -\mbox{(\# 2$^{\mbox{\tiny nd}}$ class const.)} 
     = 4\,.
\end{array}
\end{equation}
These four degrees of freedom correspond to the propagation of the graviton in the phase space.

Now we move to the side of Einstein's general relativity. Its Hamiltonian can be taken from Eq.~(\ref{hamiltonian}) by putting $\lambda = 1$ in the metric $\mathcal{G}_{ijkl}$ given in Eq.~(\ref{inverseg}). All the Eqs.~(\ref{hamiltonian}) - (\ref{brackethh}) still hold with this special value for $\lambda$. In particular, as it is well known, Eq.~(\ref{brackethh}) entails the fact that the Hamiltonian constraint $\mathcal{H}$ is a first-class constraint. Dirac's algorithm ends with Eq.~(\ref{brackethh}) since the time preservation of all the constraints is ensured.

The above Hamiltonian analysis for general relativity is gauge invariant. Since all constraints are of first-class, certain gauge conditions on the canonical variables $g_{ij}$ and $\pi^{ij}$ and the Lagrange multipliers $N$ and $N_i$ can be imposed. One partial gauge fixing condition commonly used is precisely the Eq.~(\ref{picero}), whose preservation in time, determined by the Hamiltonian of general relativity, leads to the Eq.~(\ref{nequation}) (see Ref.~\cite{DeWitt:1967yk}).

Therefore, the equivalence of the $\lambda R$-model and general relativity becomes clear. For the former $\mathcal{H}$ and $\pi$ are second class constraints and Eq.~(\ref{nequation}), which is an equation for $N$, is needed to preserve the constraints in time. On the side of general relativity, $\mathcal{H}$ is a first-class constraint, the vanishing of $\pi$ can be imposed as a gauge fixing condition and again the Eq.~(\ref{nequation}) is needed to preserve this gauge in time. All the equations involved in the Hamiltonian analysis for both theories are the same since the constant $\lambda$ drops out of the Hamiltonian due to the constraint $\pi$.

The equivalence between certain theories with first/second-class constraints is well known in the literature of constrained dynamics. First, consider a theory with a single first-class constraint $\phi$ and let $\chi$ be a chosen gauge fixing condition for the associated gauge symmetry. The appropriated measure in the path integral is given by
\begin{equation}
d\mu =
  \delta \phi \delta \chi \det \{ \phi , \chi \} \,.
\label{measure1}
\end{equation}
This measure guarantees that the path integral is independent of the chosen gauge fixing condition $\chi$. Second, the measure of a theory with two second-class constraints $\phi_1$ and $\phi_2$ is
\begin{equation}
 d\tilde{\mu} = \delta \phi_1 \delta \phi_2 \sqrt{\det \{ \phi_i , \phi_j \}} \,.
\label{measure2}
\end{equation}
Now consider the special case in which at least one of the second class constraints, say $\phi_1$, commutes with itself in the constrained phase space, $\{ \phi_1 , \phi_1 \} \cong  0$. Since in this case $\det \{ \phi_i , \phi_j \} = (\det\{ \phi_1 , \phi_2 \})^2$, the measure (\ref{measure2}) takes the form
\begin{equation}
 d\tilde{\mu} = \delta \phi_1 \delta \phi_2 \det \{ \phi_1 , \phi_2 \} \,.
\label{measure3}
\end{equation}
This measure becomes completely equivalent to (\ref{measure1}) if we reinterpret the second-class constraint $\phi_2$ as the gauge fixing condition $\chi$ for $\phi_1$, which now is interpreted as the only (first-class) constraint of the theory. The original Hamiltonian must be changed by the addition of a term which vanishes in the chosen gauge. This step is required in order to have a weakly vanishing bracket between the first-class constraint $\phi_1$ and the Hamiltonian. In the $\lambda R$-model, $\mathcal{H}$ plays the role of $\phi_1$ and $\pi$ is $\phi_2$. The Hamiltonian of general relativity is recovered when a term proportional to $(\pi)^2$ is added.

%%%%%%%%%%%%%%%%%%%%%%%%%%%%%%%%%%%%%%%%%%%%%%%%%%%%%%%%%%%%%%%%%%%%%%
%%%%%%%%%%%%%%%%%%%%%%%%%%%%%%%%%%%%%%%%%%%%%%%%%%%%%%%%%%%%%%%%%%%%%%
%%%%%%%%%%%%%%%%%%%%%%%%%%%%%%%%%%%%%%%%%%%%%%%%%%%%%%%%%%%%%%%%%%%%%%
%%%%%%%%%%%%%%%%%%%%%%%%%%%%%%%%%%%%%%%%%%%%%%%%%%%%%%%%%%%%%%%%%%%%%%
%%%%%%%%%%%%%%%%%%%%%%%%%%%%%%%%%%%%%%%%%%%%%%%%%%%%%%%%%%%%%%%%%%%%%%

\paragraph{Hamiltonian analysis for $\lambda = {1}/{3}$ \\}
As we have already mentioned, for the case $\lambda = 1/3$ \footnote{A perturbative analysis for this special case has been done in Ref.~\cite{Park:2009hg}.} the metric $G^{ijkl}$ is no longer invertible, being the metric $g_{ij}$ a null eigenvector of it, ${G}^{ijkl} g_{kl} = 0$. Equation (\ref{pi}) implies that the velocities $\dot{g}_{ij}$ can not be completely solved in terms of the momenta $\pi^{ij}$. Indeed, from Eq.~(\ref{pi}) one gets automatically a primary constraint,
\begin{equation}
 \pi = 0 \,,
\end{equation}
which was obtained in the $\lambda \neq 1/3$ case as a secondary constraint. Again, since conformal transformations are not symmetries of this theory, one expects that this is a second class constraint\footnote{This is not the case of the pure $z=3$ Ho\v{r}ava theory with the special value $\lambda = 1/3$, since this theory  possesses conformal transformations as part of its gauge symmetries. This conformal symmetry is broken by the $z=2,1$ deformations for any $\lambda$.}

Although in this case we cannot solve the velocities explicitly, it is not difficult to perform the Legendre transformation to obtain the Hamiltonian. Indeed, it is straightforward to obtain from Eq.~(\ref{pi})
\begin{eqnarray}
 \frac{\pi^{ij} \pi_{ij}}{g} & = &
   G^{ijkl} K_{ij} K_{kl} \,,
\\
\pi^{ij} \dot{g}_{ij} & = &
  \frac{2 N}{\sqrt{g}} \pi_{ij} \pi^{ij} + 2 \pi^{ij} \nabla_i N_j \,.
\end{eqnarray}
Using these relations we build the Hamiltonian and obtain that it is given exactly by Eqs.~(\ref{hamiltonian}), once the term proportional to $\pi^2$ has been dropped out from those equations. Therefore, for the $\lambda R$-model, the case of $\lambda = 1/3$ is not different to the other values of $\lambda$. The distinction lies only in the way one gets the Hamiltonian and the constraints, which at the end are all the same.

\begin{center}
{* \hspace*{5em} * \hspace*{5em} *}
\end{center}

As concluding remarks, we would like to point out some results we have obtained. In the first place, the $\lambda R$-model is consistent in describing the dynamics of two propagating degrees of freedom as general relativity. Indeed, the dynamics of both theories is exactly the same, as can be seen in a particular gauge for general relativity. Second, we emphasize the crucial point of our approach: the condition (\ref{preconstraint}) must be regarded as a second-class constraint and not as an equation for the lapse function, since Eq.~(\ref{preconstraint}) does not give any information about it. Moreover, this constraint is equivalent to the very simple constraint (\ref{picero}). Consequently, one has to ensure the Dirac conservation condition for this constraint and it happens that the procedure ends at this step since the lapse function is determined by the Eq.~(\ref{nequation}). Since the $\lambda R$-model is an effective model for large distances of the Ho\v{r}ava theory, our result gives evidence in favour of the (theoretical) admissibility of the Ho\v{r}ava Theory of quantum gravity.

Let us mention that, as in any gauge theory, there is an infinite set of admissible gauge fixing conditions for general relativity and in each case the resulting theory describes the same classical physics\footnote{The only possible obstruction could arise from gauge anomalies in the corresponding quantum field theory, an issue that is of course beyond the classical field analysis we have done.}. This kind of equivalence also holds in some cases starting with a theory with second-class constraints. We have pointed out (Eqs.~(\ref{measure1})-(\ref{measure3})) how a theory with two second-class constraints becomes equivalent to a theory with a single first-class constraint and a gauge fixing condition. The theory with the first-class constraint has an underlying gauge symmetry that is not present in the theory with the second-class constraints. This fact is already known and has been used in the literature of constrained systems.

Our results agree with the analysis of Ref.~\cite{Afshordi:2009tt} and with the perturbative analyzes done in Refs.~\cite{Gong:2010xp,Park:2009hg} for this model. Moreover, in Refs.~\cite{Gong:2010xp,Park:2009hg} higher derivative terms were included in the potential, in particular the square-Cotton term characteristic of the Ho\v{r}ava theory. Therefore, these results are encouraging to reanalyze with exact, non-perturbative computations the complete Ho\v{r}ava Theory following the mechanism we have used. Of course, the behavior at short distances, where the higher-derivative terms in the potential become dominant, could be different from that at large distances. In particular, the inclusion of the higher-derivative terms could introduce modifications to the way in which the condition (\ref{preconstraint}) is consistently solved and probably those models are no longer physically equivalent to theories that are covariant under general spacetime relativity. This includes the presence of an extra degree of freedom in the side of Ho\v{r}ava models. Such extra mode is not in conflict with the result we have obtained here whenever there exists a mechanism under which it decouples at large distances. This is a delicate point that deserves a careful analysis, since the problem of strong coupling at the IR of the extra modes have been raised by several authors, see \cite{Blas:2009yd,Charmousis:2009tc}. We expect to report on this shortly.

The form of the condition (\ref{preconstraint}) could also be modified by the inclusion of external sources. For generic matter sources the analysis of this equation changes drastically, since its explicit dependence on $N$ cannot be removed. The right hand side of Eq.~(\ref{preconstraint}) becomes nonzero and the steps be performed after it do not hold. In this case the analysis is similar to the one of Ref.~\cite{Bellorin:2010te}, where $N$ is considered as part of the canonical variables. If the closure of the algebra of constraints of the theory with sources can be proven, then an odd extra mode will arise. Therefore, the theory cannot be equated with a relativistic theory.

The Hamiltonian analysis we have done for the $\lambda R$-model can be also done for the large-distance effective action of the extended theory proposed in Ref.~\cite{Blas:2009qj} (see Ref.~\cite{Kluson:2010nf}), where spatial derivatives of the lapse function are included in the Lagrangian through extra terms that are covariant under the gauge symmetry group of the Ho\v{r}ava Theory. The large-distance truncation of this model has a term of second order in spatial derivatives of $N$, hence it does not lead to pure general relativity at large distances, in contrast with the $\lambda R$-model. However, one could be in principle in favour of adding the extra terms proposed in Ref.~\cite{Blas:2009qj} due to the renormalizability of the theory. With this regard we would like to comment that the problem of renormalization of all these theories is far from being well-defined. The main point is that they present an involved structure of second-class constraints. It is only after the understanding on the closure of these constraints and how to handle them explicitly in the quantum theory that one may know exactly which extensions should lead to a renormalizable theory. It is known that second class constraints may introduce non-localities that may ruin the quantization procedure.

%%%%%%%%%%%%%%%%%%%%%%%%%%%%%%%%%%%%%%%%%%%%%%%%%%%%%%%%%%%%%%%%%%%%%%
%%%%%%%%%%%%%%%%%%%%%%%%%%%%%%%%%%%%%%%%%%%%%%%%%%%%%%%%%%%%%%%%%%%%%%
%%%%%%%%%%%%%%%%%%%%%%%%%%%%%%%%%%%%%%%%%%%%%%%%%%%%%%%%%%%%%%%%%%%%%%
%%%%%%%%%%%%%%%%%%%%%%%%%%%%%%%%%%%%%%%%%%%%%%%%%%%%%%%%%%%%%%%%%%%%%%

\section*{Acknowledgments}
We would like to thanks  J. M. Pons and P. Talavera for their very useful comments.

%%%%%%%%%%%%%%%%%%%%%%%%%%%%%%%%%%%%%%%%%%%%%%%%%%%%%%%%%%%%%%%%%%%%%%
%%%%%%%%%%%%%%%%%%%%%%%%%%%%%%%%%%%%%%%%%%%%%%%%%%%%%%%%%%%%%%%%%%%%%%
%%%%%%%%%%%%%%%%%%%%%%%%%%%%%%%%%%%%%%%%%%%%%%%%%%%%%%%%%%%%%%%%%%%%%%
%%%%%%%%%%%%%%%%%%%%%%%%%%%%%%%%%%%%%%%%%%%%%%%%%%%%%%%%%%%%%%%%%%%%%%

\end{document}